\documentclass[twocolumn,showpacs,preprintnumbers,superscriptaddress,amsmath,floatfix,amssymb,prd]{revtex4}
\usepackage{graphicx}
\usepackage[colorlinks=true]{hyperref}
\usepackage{amsfonts,amsmath,amsxtra}

\newcommand{\beq}[1]{\begin{eqnarray}\label{#1}}
\newcommand\eeq {\end{eqnarray}}
\newcommand\bqa {\begin{eqnarray}}
\newcommand\eqa {\end{eqnarray}}
\newcommand\pr {\partial}

\newcommand{\eq}[1]{(\ref{#1})}
\newcommand{\bear}{\begin{array}}
\newcommand{\enar}{\end{array}}

\newcommand{\logo}{\\ \vskip -18mm
\leftline{\includegraphics[scale=0.3,clip=false]{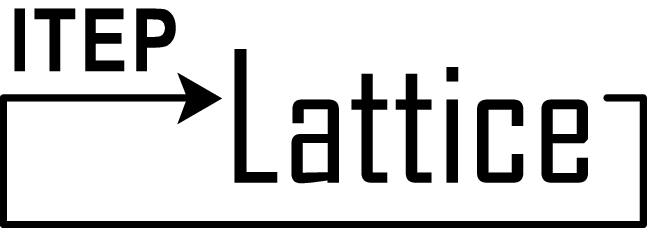}} \vskip 10mm}

\begin{document}
\sloppy \preprint{ITEP--LAT/2008--16}

\def\t{\theta}
\def\T{\Theta}
\def\w{\omega}
\def\ov{\overline}
\def\a{\alpha}
\def\b{\beta}
\def\g{\gamma}
\def\s{\sigma}
\def\l{\lambda}
\def\wt{\widetilde}


\title{Interacting Field Theories in de Sitter Space are Non--Unitary \logo}
\author{Emil T. Akhmedov}
\email{akhmedov@itep.ru}
\affiliation{ITEP, 117218 Russia, Moscow, B. Cheremushkinskaya str. 25}
\author{P. V. Buividovich}
\email{buividovich@tut.by}
\affiliation{JIPNR, National Academy of Science, 220109 Belarus, Minsk, Acad. Krasin str. 99}
\affiliation{ITEP, 117218 Russia, Moscow, B. Cheremushkinskaya str. 25}
\date{August 30, 2008}
\begin{abstract}
It is well known that there should be a total cancellation of the
IR divergences in unitary interacting field theories, such as QED
and gravity. The cancellation should be at all orders between loop
and tree level contributions to cross--sections. This is the
crucial fact related to the unitarity of the evolution operator
(S--matrix) of the underlying interacting field theory. In this
note we show that such a cancellation does {\it not} happen in de
Sitter space.
\end{abstract}
\maketitle

\section{Introduction}

It is commonly accepted that de Sitter
background should correspond to the lowest energy state in the
gravity theory with the positive cosmological constant. The main
argument behind this point of view is that de Sitter space has
the largest possible isometry group with the given cosmological
constant, while any deviation from de Sitter background breaks the
isometry. As well it can be shown that there are no exponentially
growing linearized fluctuations in de Sitter space
\cite{Lifshitz:1945du}.

Let us address, however, the following question in the {\it
interacting} field theory on de Sitter background: Does an
inertially moving charged particle in de Sitter space emit
radiation or not? Because the space in question is conformaly
flat, we propose to consider field theory which is not conformal,
otherwise the behavior of fields is not much different from the
one in Minkowski space.  For example, free electromagnetic fields
do not feel the expansion of de Sitter space and behave as if they
are in Minkowski one. However, we can consider either minimally
coupled scalars and gravitons (whose free field theories already
are non--conformal) or turn on interactions which break conformal
invariance.

It is possible to fetch the answer on the posed question even
before going into the calculational details using just general
physical arguments. In fact, an inertially moving particle in de
Sitter space accelerates with respect to a free floating
(inertial) observer in the same space. Hence, it is tempting to
think that inertial particle should radiate from the point of view
of the observer in question. We support these general comments
with the explicit calculation in the main body of the text.
Somewhat similar phenomenon have been considered in
\cite{Myrhvold}.

But if the particle radiates it will do that eternally --- as long
as the particle and the background are left untouched. Where does
the energy for the radiation come from? One can object that we
should not ask about energy in a time dependent de Sitter like
background. But exactly in this objection resides the answer to our
question. In fact, the Hamiltonian of an {\it interacting
non--conformal} field theory in de Sitter background does not have
a ground state, if the cosmological constant is held fixed. We
argue that the radiation happens at the cost of the decrease of
the cosmological constant, if the latter is not supported to be
constant by any imaginary external influence.

What we are trying to point out here is that the situation in many
respects is similar to the one in QED with the constant electric
field if we use the creation/annihilation operators which
correspond to the exact harmonics rather than just to plain waves.
As is well known the similarity goes even further --- up to the
pair creation \cite{Gibbons:1977mu} --- but to avoid any mystery
of quantum field theory in curved backgrounds with horizons, we
prefer to discuss a simple semiclassical (tree level QFT)
phenomenon and IR rather than UV behavior of quantum corrections,
which can be completely understood with the presently existing
level of knowledge. It is obvious that, even if we neglect the
pair production, a charge placed in the constant electric field
background will radiate at the cost of the decrease of the field
in question or at the cost of work performed by the external
source keeping the electric field constant. Similarly the
cosmological constant should decrease if it is not held fixed by
any external source.

In this note we would like to find a straightforward signal
showing that there are problems with quantum fields in de Sitter
background. Because of asymptotic non--flatness of de Sitter space
(hence, no energy conservation) the radiation discussed above
could be considered as not being a problem, but we show that it
inevitably results in one. We observe that the evolution operator
in de Sitter space is not unitary if we keep the cosmological
constant fixed. The way to see that is through the
non--cancellation of the IR divergences between tree and loop
contributions to the cross--sections.

Let us sketch here the arguments presented in the main body of the
text. The statement is that in de Sitter space a charged particle
on mass--shell does emit radiation. Hence, its virtuality
\footnote{E.g. in a QFT without background field the virtuality
$\lambda$ is just the difference between the energy squared and
the momentum squared of the particle ($\lambda = k_0^2 -
\vec{k}^2$), i.e. the measure of its off--shell position.} is not
related to the momentum of the emitted radiation. Thus, the tree
level cross--sections for emmitting soft radiation are finite. In
fact, recall that IR divergences in cross--sections (of a QFT
without background fields) appear because the propagators of the
particles which create radiation, being proportional to their
inverse virtuality, are singular as the momentum of the emitted
radiation goes to zero \cite{Weinberg:1965nx}: As follows from the
energy--momentum conservation the virtuality is proportional to
the momentum of the soft radiated quantum. The power of the
singularity is such that, after the integration over the phase
volume of the radiated soft quantum in any cross--section, the
logarithmic IR divergence appears. Such IR divergences cancel with
the ones appearing in loop corrections to the cross--sections
\cite{Weinberg:1965nx}. The cancellation can be directly linked to
the unitarity of the S--matrix in the QFT \cite{Lee:1964is} or,
more concretely, to the optical theorem.

Now in the case of de Sitter space, while tree level
cross--sections are finite, the loop diagrams do have IR
divergences! As a result, unlike the situation in a QFT without
background fields, there is nothing which can cancel them. Hence,
the evolution operator in de Sitter space is not unitary, as it
should be for a non--closed system due to the presence of a
background field which is held fixed by an external source.

It is worth pointing out here that the problems with cancellation
of the IR divergences appear in de Sitter space even if we respect
the de Sitter isometry at every step of the calculation, i.e. de
Sitter space is unstable and the isometry is broken, at least if
one turns on interactions. It is unstable in the sense that the
cosmological constant will decrease, which will result eventually
in FRW universe (with non--accelerating expansion). In the latter
case we are going to have an analog of the standard Minkowski
vacuum for quantum fields.

\section{General discussion}

We set the cosmological constant to be one and keep it fixed
throughout the paper. Although our arguments are general, for
simplicity we would like to consider two minimally coupled real
scalar particles in $D$--dimensional de Sitter space with the
Yukawa type interaction:

\begin{widetext}
\bqa S_{matter} = \frac12\,\int d^Dx \, \sqrt{-g} \, \left[g^{ab}
\, \pr_a \Psi \pr_b \Psi + M^2 \, \Psi^2 + g^{ab} \, \pr_a \psi
\pr_b \psi + m^2 \, \psi^2 + \lambda\, \Psi^2 \, \psi
\right]\label{theory}\eqa
\end{widetext} It will become clear from the discussion below that
the reason for consideration of such a theory is that we would
like to keep both masses $m$ and $M$ greater than zero.

This theory in Minkowski space does possess the cancellation of
the IR divergences if $m=0$ and $M>0$ or does not have them at all
if $M, m>0$. As well one can consider non--minimally coupled
scalars if she will make such substitutions as $m^2 \to m^2 +
\zeta \, R$ for all scalar masses, where $R$ is the de Sitter
curvature, and $\zeta$ is a parameter. Conformal coupling
corresponds to the case $m=0$, $\zeta = (D-2)/4(D-1)$. It is
important that the interaction term breaks the conformal
invariance in any case.

At first sight the most convenient reference frame where one can
do all the calculations is the planar one \bqa ds^2 = - dt^2 +
e^{2\,t}\, dx_i \, dx_i = \frac{1}{\tau^2}\, \left(- d\tau^2 +
dx_i \, dx_i\right)\label{planar}\eqa where $\tau = e^{-t}$,
because then we have to deal with the non--compact spacial
sections and the formulas for amplitudes are very similar to those
in Minkowski space QED with constant electric field. Unfortunately
in this coordinates we encounter problems. To see them consider
the Klein--Gordon equation describing propagation of free waves in
these coordinates:

\bqa \left(\tau^2 \, \pr_{\tau}^2 - (D-2)\, \tau\, \pr_{\tau} -
\tau^2 \, \pr_i \, \pr_i + m^2\right) \psi = 0\label{waveplan}\eqa
and similarly for $\Psi$. Because this is a free wave equation,
its solutions obey the superposition principle. Hence, from the
point of view of the observer, seeing the corresponding metric,
particles (single waves) are just solutions of such an equation,
having a finite flux to be defined below. But we would like to
respect the de Sitter isometry, which restricts the choice of the
basis of harmonics. A particular basis which leads to the de
Sitter invariant vacuum state is as follows \cite{Bunch:1978yq}:

\bqa \psi_k \propto e^{i \, \vec{k}\, \vec{x}} \, \tau^{(D-1)/2}
H^{(2)}_\nu (k\tau)
\nonumber \\
\nu = \sqrt{\left(\frac{D-1}{2}\right)^2 - m^2}, \quad k =
|\vec{k}| \label{Hank}\eqa These harmonics correspond to the
positive energy states, while their complex conjugates --- to the
negative. Here $H^{(2)}$ is the Hankel function: $H^{(1)}_\nu =
\left(H^{(2)}_\nu\right)^*$.

Let us stress here the main problem with these harmonics. The term
linear in the differential over $\tau$ under the brackets in
\eq{waveplan}, which we refer to as a ``friction'' term, has a
wrong sign as $\tau$ goes to $+\infty$ (past infinity). As the
result among the harmonics present in the complete basis of the
solutions of this equation we have those which are exponentially
big (in $t$) in the past, when $D\geq 3$. In fact, the solution
presented in \eq{Hank} behaves, when $\tau\to +\infty$ ($t\to
-\infty$), as follows:

\bqa \psi_k \propto \tau^{\frac{D-2}{2}} \, e^{- i \, k \, \tau +
i \, \vec{k}\, \vec{x}} \label{behavior}\eqa This happens because
the metric in \eq{planar} is singular as $\tau\to +\infty$. As the
result all loop diagrams have divergences in the $\tau \to
+\infty$ corner of the time integration axis. Such a divergence
present along with the IR one which appears at $\tau = 0$ corner.

We postpone the discussion of the problems with the definition of
the S--matrix in de Sitter space to the following sections and
define here the mass--shell three leg amplitude as follows. It is
proportional to the integral:

\begin{widetext}
\bqa A \propto \left\langle k, q\left| \int d^D x \, \sqrt{-g} \,
\hat{\Psi}^2 \, \hat{\psi} \right| p\right\rangle \propto \nonumber \\
\propto \int_0^{+\infty} \frac{d\tau}{\tau^D} \,\int
d^{(D-1)}\vec{x} \, e^{i \, \left(\vec{p} - \vec{k} -
\vec{q}\right)}\, \tau^{\frac{3}{2}\, (D-1)} \,
H^{(1)}_{\nu_1}(p\tau)\, H^{(2)}_{\nu_1}(k\tau) \,
H^{(2)}_{\nu_2}(q\tau) = \nonumber \\ =
\delta^{(D-1)}\left(\vec{p} - \vec{k} - \vec{q}\right) \,
\int_0^{+\infty} d\tau \tau^{\frac{D-3}{2}}\,
H^{(2)}_{\nu_1}(p\tau)\, H^{(1)}_{\nu_1}(k\tau) \,
H^{(1)}_{\nu_2}(q\tau)\label{amplitude}\eqa
\end{widetext}
Here $|k,q\rangle = \hat{a}^+_k\, \hat{b}^+_q\,|vac\rangle$ and
etc.. Here $\hat{a}$ and $\hat{b}$ are creation operators for the
harmonics \eq{Hank} of the fields $\Psi$ and $\psi$,
correspondingly; $|vac\rangle$ is the de Sitter invariant
Bunch--Davies vacuum \cite{Bunch:1978yq} and $\nu_1 =
\sqrt{\left(\frac{D-1}{2}\right)^2 - M^2}, \quad \nu_2 =
\sqrt{\left(\frac{D-1}{2}\right)^2 - m^2}$. Because of such a
behavior as in \eq{behavior}, the integral in \eq{amplitude} looks
like:

$$A \propto \int^{+\infty} d\tau \, \tau^{\frac{D-6}{2}} \, e^{i\,(p-k-q)\,
\tau}$$ at the upper integration limit. Hence, for $D\geq 5$ we
have a divergence even in the tree level amplitude independently
of the value of $m$ and $M$. Similar problems appear in the loop
diagrams starting with $D = 4$.

 One can try to avoid the divergence by
turning on the interactions at some finite $\tau_0$ and evolve to
a future $\tau < \tau_0$. This is explicitly done in loop
amplitudes in the papers \cite{Tsamis:1996qq} and implicitly ---
in the papers \cite{Dolgov:1994cq} and \cite{Weinberg:2006ac}.
However, in this way we break the de Sitter invariance by hand,
because the latter acts on $\tau_0$. Hence, it is not an occasion
that in the quoted papers a perturbation of the de Sitter metric
which does not respect the invariance was observed. It is not that
we completely disagree with such an approach, taking into account
that de Sitter invariance is going to be dynamically broken
anyway, but we just would like to show here that one will
encounter problems even if she will always try to respect the
invariance.

Before going further let us point out the meaning of the amplitude
\eq{amplitude}. First, let us stress that the calculation of the
amplitude gives a generally covariant (and gauge invariant) way to
address the question of radiation. In fact, if it is not vanishing
for given directions of external momenta and when all its external
legs are on mass--shell, it just means that a single wave can
split into two waves from the point of view of the observer
corresponding to the background metric, in which all calculations
have been done.

Second, notice that the amplitude \eq{amplitude} is proportional
to the spacial $\delta$--function imposing the momentum
conservation law $\vec{p} = \vec{k} + \vec{q}$. In Minkowski space
the time integral would as well impose the energy conservation law
$p_0 = k_0 + q_0$ via $\delta$--function. At the same time on
mass--shell in Minkowski space we obtain $p_0 = \sqrt{\vec{p}^2 +
M^2}$, $k_0 = \sqrt{\vec{k}^2 + M^2}$, $q_0 = \sqrt{\vec{q}^2 +
m^2}$ and the energy conservation condition does not have a
solution. Hence, the amplitude is zero and mass--shell (inertial)
particles in Minkowski space can not emit radiation.

Now, in \eq{amplitude} we do not have the energy conservation due
to the presence of the external gravitational field originating
from the cosmological constant --- similarly to the QED in an
external constant electric field. Thus, the amplitude is not zero
on mass--shell. For the case when it is convergent (i.e. when
$D\leq 4$ and $M,m> 0$) this can be seen explicitly via numerical
calculation of the integral in \eq{amplitude} using Mathematica or
Maple.

In our opinion, the aforementioned divergencies in loop diagrams
at the past infinity ($\tau=+\infty$) are simply sort of boundary
effects, which emerged because the planar coordinates cover only
half of de Sitter space. To avoid the aforementioned divergencies
in loop diagrams at the past infinity ($\tau=+\infty$) we propose
to consider the global coordinate system:

\bqa ds^2 = - dt^2 + \cosh^2 t \,
d\Omega^2_{D-1}\label{global}\eqa where $d\Omega^2_{D-1} =
d\theta^2_1 + \sin^2 \theta_1 \, d\theta^2_2 + \dots + \sin^2
\theta_1 \dots \sin^2\theta_{D-2} \, d\theta^2_{D-1}.$ Unlike the
planar coordinates, the global ones cover de Sitter space
completely. Important feature of these and the planar coordinates
is that they are seen by inertial observers. The Klein--Gordon
equation in these coordinates is as follows:

\bqa \left(\pr_t^2 + (D-2) \tanh t \, \pr_t + m^2 -
\frac{\Delta_{D-1}(\Omega)}{\cosh^2 t}\right) \, \psi =
0\label{waveglobal}\eqa Here $\Delta_{D-1}(\Omega)$ is the
Laplacian on the $(D-1)$--dimensional sphere.

The ``friction'' term in \eq{waveglobal} is proportional to $\tanh
t$ and changes sign from $-(D-1)$ in the past infinity ($t\to
-\infty$) to $+(D-1)$ in the future infinity ($t\to + \infty$). As
the result we obtain complete set of modes which are finite at
every value of $t$ at any given mass $m$. To solve \eq{waveglobal}
explicitly we can use the separation of variables:

\bqa
 \psi_{j\, n}(t, \Omega) = \varphi_j(t) \, Y_{j \, n}(\Omega)\label{jmharm}
\eqa Here $\Delta_{D-1}(\Omega) \, Y_{j\, n}(\Omega) = - j(j +
D-2)\, Y_{j\, n}(\Omega)$, and $n$ is the multi--index
$(n_1,\dots, n_{D-2})$.

The spherical harmonics
$Y_{j\, n}(\Omega)$ have obvious properties presented in
\cite{Bousso:2001mw}. The field $\varphi_j(t)$ obeys the obvious
equation following from \eq{waveglobal} and \eq{jmharm}. This
equation has two designated complete sets of solutions: so called
``in'' and ``out'' modes \cite{Mottola:1984ar} (see as well
\cite{Bousso:2001mw}). The complete set of ``in'' modes is

\begin{widetext}
\bqa \varphi_j^{\pm}(t) \propto \cosh^j(t) \, e^{\left(j +
\frac{D-1}{2} \mp i \, \mu\right)\, t} \, F\left(j +
\frac{D-1}{2}, \, j + \frac{D-1}{2} \mp i\,\mu; 1 \mp i\, \mu; -
e^{2\,t} \right)\label{solution}\eqa
\end{widetext}
where $ \mu = \sqrt{m^2 - \left(\frac{D-1}{2}\right)^2}$ and $F(a,b;c;z)$ is the
hypergeometric function. The solution \eq{solution} can be
continued to the case when $m < (D-1)/2$. The ``out'' modes
$\bar{\varphi}^{\pm}_j(t)$ are related to the ``in'' modes as
follows $\bar{\varphi}^{\pm}_j(t) = \left(\varphi^{\pm}_j(-
t)\right)^*.$

The ``in'' or ``out'' wave functions in question (as well as
\eq{Hank}) are orthonormal with respect to the norm $i\, \int_X \,
\psi_1 \sqrt{-g}\, g^{00}\, \pr_0 \, \psi^*_2 \, d^{D-1} \Omega $,
which is invariant under the change of the spacial section $X$, as
the consequence of the equation \eq{waveglobal} (or
\eq{waveplan}). This norm defines the flux. Hence, any solution of
the eq. \eq{waveglobal} which has a definite finite flux (i.e.
corresponds to a propagating particle) can be decomposed in the
complete basis of easer ``in'' or ''out'' modes. The particular
choice of the ``in'' or ``out'' modes as the basis of harmonics
leads to the de Sitter invariant vacuum state
\cite{Mottola:1984ar}.

For the future references let us discuss here the asymptotic
behavior of the ``in'' modes.  The hypergeometric function
$F(a,b;c;z)$ does not have any poles on the negative $z$ axis
($z=-e^{2\,t}$ in our case). Hence, the ``in'' modes \eq{solution}
are regular at any value of $t$ and $m$ and behave in the past
infinity ($t\to-\infty$) as

\bqa \varphi_j^{\pm} \rightarrow e^{\left(\frac{D-1}{2} \mp i \,
\mu\right)\, t}\label{tmininf}\eqa because $F\to 1$ as $z=
-e^{2t}\to 0$. In the future infinity ($t\to +\infty, \,
z=-e^{2t}\to -\infty$) they look like

\bqa \varphi_j^{\pm} \rightarrow e^{-\frac{D-1}{2}\,t}\left(c_1 \,
e^{\mp i \,\mu \, t} + c_2 \, e^{\pm i \,\mu \, t}\right)
\label{tpluinf}\eqa with some complex constants $c_1$ and $c_2$.
Such a behavior follows from

\bqa \lim_{z\to-\infty} F(a,b;c;z) \rightarrow c_1 \, (-z)^{-a} +
c_2 \, (-z)^{-b}\label{propasymp}\eqa Note that if $m=0$ we have
harmonics which approach non--vanishing constants as $t\to
\pm\infty$, but there are no any modes which are exponentially
growing.

For the future reference let us define here the propagator for the
``in'' modes \cite{Mottola:1984ar}, \cite{Bousso:2001mw}. Consider
de Sitter invariant function depending on two points
\cite{Mottola:1984ar}, \cite{Bousso:2001mw}:

\bqa Z(z,z') = - \sinh t\, \sinh t' + \cosh t \, \cosh t' \,
\cos\Delta\Omega\label{subst}\eqa
where $\Delta\Omega$ is the
angle between the spacial parts of the coordinates $z$ and $z'$.
It can be shown that $Z(z,z') = \cos L,$ where $L$ is the geodesic
distance between $z$ and $z'$ for spacial separations, or $i$
times the geodesic proper time difference for time--like
separations. The Green function, being de Sitter invariant, should
depend only on such a combination of the two points. Hence, the
equation for the Green function, which is \eq{waveglobal} with the
appropriate $\delta$--functional sources on the RHS, can be
converted into:

\bqa \left[\left(1-Z^2\right)\, \pr_Z^2 - D\, Z\, \pr_Z - m^2
\right] G(Z)
 = \nonumber \\ =
A \, \delta(Z+1) + B\delta(Z-1)\eqa by the direct change of
variables \eq{subst}, from $(t,\Omega)$ to $Z$. Here $A$ and $B$
are some constants. The RHS of this equation is singular when the
$z$ and $z'$ points coincide (i.e. when $Z=1$) and when $z$
coincides with the antipodal point of $z'$  (i.e. when $Z=-1$)
\cite{Mottola:1984ar}. The ``in'' Feynman type propagator obeys
this equation with $A=B=1$ and is given by \cite{Mottola:1984ar}:
\begin{widetext}
\bqa G_{in}(Z) \propto \frac{\mu\, (\mu + 1)}{\sinh (\pi\, \mu)}
\, \left[F\left(\frac{D-1}{2} + i\, \mu, \frac{D-1}{2} - i\, \mu;
\frac{D}{2}; \frac{1 + Z}{2} \right)  \right. + \nonumber \\ +
\left. F\left(\frac{D-1}{2} + i\, \mu, \frac{D-1}{2} - i\, \mu;
\frac{D}{2}; \frac{1 - Z}{2} \right)\right]\label{greenfunc}\eqa
\end{widetext}
where $\mu$ is defined above. We are going to use this function in
the loop calculations below.

\section{On QFT with compact spacial sections}

 We see that spacial sections in global coordinates are compact $(D-1)$--dimensional
spheres, which is less convenient than flat sections in planar
coordinates. To see that compactness of spacial sections does not
spoil all the picture, in this chapter we would like to consider a
general features of QFT on such a space--time as, for example,
$ds^2 = - dt^2 + R^2 \, d\Omega^2_{D-1}$ with fixed radius $R$. We
are going to show that such a QFT as \eq{theory} on the background
in question has similar properties to the theory in Minkowski
space: such properties as the impossibility for inertial particle
to emit radiation as measured by an inertial observer and the
cancellation of IR divergences. Let us stress here that our
confidence in the fact that there are no problems with the theory
\eq{theory} on the background in question is relaying on the
obvious observation that it has the unitary evolution operator.

Definite energy mass--shell harmonics in such a theory look like $
\phi_j \propto e^{-i\,k_0 \,t}\, Y_{j\, m}(\Omega)$, where $k_0 =
\sqrt{m^2 + \frac{1}{R^2} \, j \, (j + D-2)}$. Then, the three leg
mass--shell amplitude is:

\bqa A \propto \int_{-\infty}^{+\infty} dt \, e^{- i\,(p_0 - k_0 -
q_0)\, t}\,
\nonumber \\
 \int d\Omega Y_{j_1\,n_1}(\Omega)\,
Y^*_{j_2\,n_2}(\Omega)\, Y^*_{j_3\,n_3}(\Omega)\eqa and is
proportional to $\delta(p_0 - k_0 - q_0)$. The second integral
(over the angles) gives the generalization of the $3j$ symbols to
the $SO(D-1)$ group with $D\ge 4$, which is not quite convenient
object in comparison with the $\delta$--function appearing for the
case of the flat spacial sections. The $3j$ symbols are nonzero if
$j_2 - j_3 \leq j_1 \leq j_2 + j_3$.

On mass--shell we have that $$p_0 = \sqrt{M^2 + \frac{1}{R^2} \,
j_1 \, (j_1 + D-2)},$$ $$k_0 = \sqrt{M^2 + \frac{1}{R^2} \, j_2 \,
(j_2 + D-2)},$$ $$q_0 = \sqrt{m^2 + \frac{1}{R^2} \, j_3 \, (j_3 +
D-2)}.$$ With such $p_0$, $k_0$ and $q_0$ the condition $p_0 = k_0
+ q_0$ can not be saturated for $j_2 - j_3 \leq j_1 \leq j_2 +
j_3$. Hence, the argument of the $\delta$--function imposing the
energy conservation is always non--zero, i.e. the amplitude itself
is zero. Thus, similarly to the Minkowski space, we arrive at the
obvious conclusion that in the space in question an inertial
(mass--shell) particle can not emit radiation.

 Recall that in Minkowski space IR divergences in the
cross--section appear only when $m=0$. Amplitude of a hard process
containing emittion of one soft $\psi$ mass--shell quantum by the
heavy $\Psi$ particle has the three leg multiplicative
contribution:

\bqa \int_{-\infty}^{+\infty} dt \, \int d\Omega \,
G_M(t',\,\Omega';\,\, t,\,\Omega)
\nonumber \\
 \, e^{- i\,(p_0 - q_0)\, t}\,
Y_{j_1\,n_1}(\Omega)\, Y^*_{j_3\,n_3}(\Omega)
\label{amplitude1}\eqa where $G_M$ is the propagator of the $\Psi$
field, i.e. one of the legs in the amplitude is off--shell. The
propagator is

\bqa G(t,\, \Omega;\,\, t',\, \Omega') = \sum_\lambda
\frac{\Psi_\lambda(t,\, \Omega)\, \Psi^*_\lambda(t', \,
\Omega')}{\lambda}\label{eigenval}\eqa where $\left[\square -
M^2\right]\, \Psi_\lambda = \lambda\, \Psi_\lambda$, and $\quad
\square = - \pr_t^2 + \frac{1}{R^2} \, \Delta_{D-1}(\Omega)$.
Obviously $\Psi_\lambda \propto e^{-i\, k_0 \, t}\, Y_{j_2\,
n_2}(\Omega)$, where now $k_0 = \sqrt{M^2 + \lambda +
\frac{1}{R^2} \, j_2 \, (j_2 + D-2)}$, i.e. $\sum_\lambda = \int
dk_0 \, \sum_{j_2=0}^{+\infty}$. The integral over $t$ in
\eq{amplitude1} leads to the energy conservation of the form:

\bqa \sqrt{M^2 + \frac{1}{R^2} \, j_1 \, (j_1 + D-2)}
 = \nonumber \\ =
 \sqrt{M^2
+ \lambda + \frac{1}{R^2} \, j_2 \, (j_2 + D-2)}
 + \nonumber \\ +
 \frac{1}{R}\,
\sqrt{j_3 \, (j_3 + D-2)}\eqa For the big $j_1 \sim j_2$ and small
(soft) $j_3$ we have the solution of this equation as follows:
$\lambda \approx -2\left(p_0 \, q_0 - \frac{1}{R^2}\, j_1
\,j_3\right)$, where $p_0 = \sqrt{M^2 + \frac{1}{R^2} \, j_1 \,
(j_1 + D-2)}$ and $q_0^2 \approx \frac{1}{R^2}\,j_3\, (D-2)$.
Hence, the amplitude is divergent as $1/\lambda \propto
1/\sqrt{j_3}$ when $j_3 \to 0$, while the cross--section is
divergent as $1/j_3$. Similar divergences (with the opposite sign)
appear in one loop contributions to the cross--sections.

To clarify the situation let us consider the Minkowski space
($ds^2 = - dt^2 + d\vec{x}^2$) variant of the theory in question
and the part of an amplitude responsible for the radiation of a
soft $\psi$ particle by the hard $\Psi$ one. For simplicity here
we restrict ourselves to the four space--time dimensions. If we
insert the Minkowski space analog of the propagator \eq{eigenval}
into the amplitude, we can use the $\delta$--functions, imposing
energy--momentum conservation at the vertex, to fix $\lambda =
k_0^2 - \vec{k}^2 - M^2 = (p_0 - q_0)^2 - (\vec{p} - \vec{q})^2 -
M^2$. Here $p$ is the four--momentum of on--shell incoming $\Psi$
particle, $q$ is the four--momentum of the on--shell outgoing
radiated $\psi$ particle and $k$ is the four--momentum of the
off--shell (with virtuality $\lambda$) outgoing, after the
radiation, $\Psi$ particle.

If we consider radiation of the very soft particle, i.e. the
modulus of the corresponding $D$--momentum is $|q| \to 0$, then
$k$ is very close to the mass--shell, while $p^2 = M^2$  and $q^2
= 0$, i.e. $\lambda = - 2\, p\,q$, because $m=0$. Thus, the
propagator is singular as $|q| \to 0$. Moreover, such a dependence
of $\lambda$ on $q$ is important for the factorization of the IR
divergences in the cross--sections for radiation of many soft
quanta \cite{Smilga:1985hp}, which, in its own right, is crucial
for the total cancellation of all divergences.

As the result, after the integration of the differential
cross--section for the radiation of one soft quantum over its
invariant phase volume, we obtain:

$$ \int \left|A\right|^2 \, \frac{d^3\vec{q}}{|\vec{q}|}
\propto \int \frac{1}{(p\, q)^2}\, \frac{d^3\vec{q}}{|\vec{q}|} \propto
\log m_0.$$ This is the IR divergence with the cutoff $m_0\to 0$.
Because all IR divergences are of the same order, we have to sum
such contributions over all external legs of the hard process in
question \cite{Weinberg:1965nx}, \cite{Smilga:1985hp}. Then
similar divergences (with the opposite sign) appear in the loop
contributions to the cross--section of the hard process:

$${\rm loop \,\, IR\,\, divergence} \propto \int d^4q \frac{1}{(p\,q)^2 \, q^2}.$$
All such contributions (from loops and tree level diagrams) add
up, so that every divergence does cancel \cite{Smilga:1985hp}.
Higher loop contributions cancel with the divergences coming from
multiple soft quantum radiations \cite{Weinberg:1965nx}. It is
important to stress here that we can choose another basis of
harmonics for $\Psi$ (dressed with the cloud of $\psi$'s) such
that IR divergences will not be present neither in trees nor in
loops \cite{Lee:1964is}, but it is impossible to get rid of the
divergences only in trees without cancelling them in loops or vise
versa.

\vspace{3mm}

\section{Radiation and IR divergences in de Sitter space}

 In this section we are
going to show that in de Sitter space the IR divergences do not
cancel already at the leading order. But before going into the
calculational details let us note that we avoid using the term
S--matrix in de Sitter space.  The latter is build on the basis of
matrix elements of the evolution operator $\hat{S} = T\, e^{-i\,
\int _{-\infty}^{+\infty} dt \, H_{int}(t)}$, where $H_{int}$ is
the interaction Hamiltonian. The matrix elements in question are
defined with respect to the basis of states $\langle out| \,
\hat{a}\dots \hat{a}$ and $\hat{a}^+ \dots \hat{a}^+ \, |in
\rangle$ while it is assumed that $|out \rangle \equiv e^{- i\,
\int_{-\infty}^{+\infty} H_0 \, dt}\, |in \rangle = ({\rm
phase})\, |in \rangle$, where $H_0$ is the free Hamiltonian. In
the odd space--time dimensional de Sitter spaces $|out \rangle =
|in\rangle$. Hence, in odd dimensions we are safe and our
arguments pass smoothly \cite{Bousso:2001mw}.

But it appears that in even dimensional de Sitter spaces $|in
\rangle \neq |out \rangle$ \cite{Mottola:1984ar},
\cite{Bousso:2001mw}. In fact, in even dimensions $|out \rangle
\equiv e^{- i\, \int_{-\infty}^{+\infty} H_0 \, dt}\, |in \rangle
\neq ({\rm phase})\, |in \rangle$, because $|in \rangle$ is not an
eigen--state of the free Hamiltonian. The latter has the form
$\hat{H}_0 \propto \sum_k \left[ A(t)\,\hat{a}^+_k \, \hat{a}_k +
B(t)\, \hat{a}_{-k}\, \hat{a}_k + B^*(t)\, \hat{a}^+_{-k}\,
\hat{a}^+_k \right]$ with some functions of time $A(t)$ and $B(t)$
\cite{GribMamaevMostepanenko}.  Exactly due to the presence of the
non--diagonal terms $\hat{a}_{-k}\, \hat{a}_k$ and
$\hat{a}^+_{-k}\, \hat{a}^+_k$ in the Hamiltonian we have to make
the Bogolyubov transformation to diagonalize it and to observe the
particle production \cite{Gibbons:1977mu}.

Thus, $|out \rangle$ state differs from $|in\rangle$ by the
presence of the created particles, which can be explicitly
established by the following relation $|out \rangle =
\hat{V}(\hat{a}, \hat{a}^+) |in \rangle$ with some operator
$\hat{V}$ \cite{GribMamaevMostepanenko}. In this note we would
like to consider the matrix elements of the evolution operator
which are of the form $\langle in|\, \hat{a}\dots \hat{a}\,
\hat{S} \, \hat{a}^+\dots \hat{a}^+ \,|in \rangle$. Physically
this means that we neglect the particle production by the external
field and consider only scattering amplitudes in such a
background. We strongly believe that this is sufficient to make a
statement about (non--)unitarity of the evolution operator
$\hat{S}$ itself. To see that our arguments are meaningful one can
consider the similar situation appearing in QED in the background
of the constant electric field.

It is straightforward to find the basic tree level {\it
mass--shell} amplitude describing the process when $\Psi$ particle
radiates the $\psi$ particle. In the global coordinates the
amplitude is proportional to:

\begin{widetext}
\bqa A \propto \int d\Omega Y_{j_1\,n_1}(\Omega)\,
Y^*_{j_2\,n_2}(\Omega)\, Y^*_{j_3\,n_3}(\Omega)\,
\int_{-\infty}^{+\infty} dt \cosh^{D-1}(t) \times \nonumber \\
\times \left[\cosh^{j_1}(t) \, e^{\left(j_1 + \frac{D-1}{2} + i \,
\mu_1\right)\, t} F\left(j_1 + \frac{D-1}{2}, j_1 + \frac{D-1}{2}
+ i \, \mu_1; 1 + i\, \mu_1; -e^{2\,t} \right)\right]\times
\nonumber \\ \times \left[\cosh^{j_2}(t) \, e^{\left(j_2 +
\frac{D-1}{2} - i \, \mu_1\right)\, t} F\left(j_2 + \frac{D-1}{2},
j_2 + \frac{D-1}{2} - i \, \mu_1; 1 - i\, \mu_1; -e^{2\,t}
\right)\right]\times \nonumber \\ \times \left[\cosh^{j_3}(t) \,
e^{\left(j_3 + \frac{D-1}{2} - i \, \mu_2\right)\, t} F\left(j_3 +
\frac{D-1}{2}, j_3 + \frac{D-1}{2} - i \, \mu_2; 1 - i\, \mu_2;
-e^{2\,t} \right)\right]
\label{ampmasshel}\eqa
\end{widetext}
 where $\mu_1 = \sqrt{M^2 - \left(\frac{D-1}{2}\right)^2}$, $\mu_2 = \sqrt{m^2
- \left(\frac{D-1}{2}\right)^2}$.

 Note that \eq{ampmasshel} is valid even if $m$ or $M$ are less than
$(D-1)/2$. Such an amplitude is just an analytical continuation of
the corresponding generalized $3j$--symbol, which follows from the
continuation $S^D \to dS^D$. From this we can already argue that
the mass--shell amplitude \eq{ampmasshel} is non--zero.

However, if either one of the masses, $m$ or $M$, is vanishing,
the integral for the amplitude is divergent (see below). We
discuss the meaning of these divergences in the concluding
section. At this stage we would like to avoid such problems with
divergences of the tree level amplitudes and to keep our
discussion as transparent as it is possible for the case in
question. It happens that, if we keep both masses $M$ and $m$
non--zero, the integral \eq{ampmasshel} is convergent.
Unfortunately, even Mathematica and Maple refuse to take such an
integral analytically. Let us show explicitly that it is really
convergent.

The integrand expression in \eq{ampmasshel} can hypothetically
blowup only if $t\to\pm\infty$, because the hypergeometric
function is regular for the negative argument, i.e. for any finite
value of $t$. As $t\to -\infty$ we can use the behavior of the
``in'' harmonics from \eq{tmininf} to obtain that the integrand
expression in \eq{ampmasshel} approaches $e^{(\frac{D-1}{2} - i\,
\mu_2)\, t}$ for the lower integration limit. Hence, the integral
is convergent in this corner of the integration axis even if
$\mu_2$ is purely imaginary, i.e. when $m < (D-1)/2$. Indeed in
the latter case $|\mu_2| \le (D-1)/2$. This inequality is
saturated only when $m=0$. It is only in the latter case there can
be the perfect cancellation of the exponential suppression
$e^{(\frac{D-1}{2} - i\, \mu_2)\, t} \to 1$ as $t\to-\infty$, and
we have the divergent amplitude.

In the other corner of the integration axis, i.e. when $t\to +
\infty$, we can use the asymptotics as in \eq{tpluinf} and find
that the integrand behaves as

\bqa
e^{-\frac{D-1}{2}\, t}\, \left(c_1 \, e^{+i\,\mu_1 \, t} + c_2 \,
e^{-i\, \mu_1 \, t}\right)\,
\nonumber \\
 \left(c'_1 \, e^{-i\,\mu_1 \, t} +
c'_2 \, e^{+i\, \mu_1 \, t}\right)\, \left(c''_1 \, e^{-i\,\mu_2
\, t} + c''_2 \, e^{+i\, \mu_2 \, t}\right) \eqa Hence, in this
corner of the $t$ integration axis the integral \eq{ampmasshel} is
convergent if $\mu_1$ is real and $m\neq 0$.

\begin{figure}
  \includegraphics[width=6cm, angle=-90]{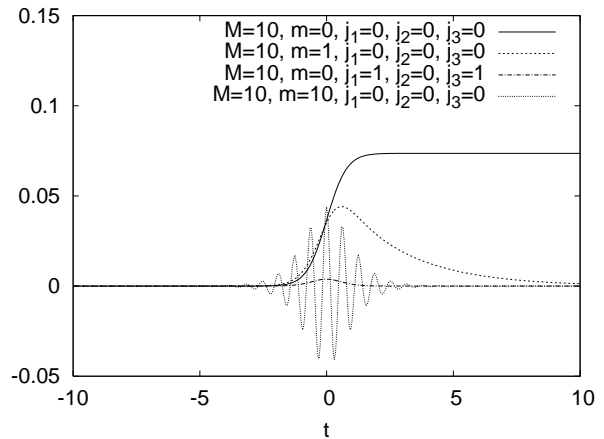}\\
  \caption{Real part of the integrand in \eq{ampmasshel} for different values of
  $M$, $m$ and $j_1$, $j_2$, $j_3$. Imaginary part is equal to zero for the first three plots.
  The presence of non--zero imaginary contribution to the leading amplitude is already a sign favoring
  that the theory in question is non--unitary.}
  \label{fig:integrand}
\end{figure}

As well, using Mathematica and Maple for numerical calculation of
the integral \eq{ampmasshel}, one can explicitly see that it is
not zero for $j_2 - j_3 \leq j_1 \leq j_2 + j_3$. The integrand
expression in \eq{ampmasshel} is plotted on Fig.
\ref{fig:integrand} for several different values of $M$, $m$ and
$j_1$, $j_2$, $j_3$. Thus, all our considerations so far at least
mean that a massive particle can radiate another massive particle
on mass--shell in de Sitter space.

Based on these considerations we can make a general conclusion
that mass--shell particles can radiate fields under which they
carry charges, unless the corresponding theory, describing
interactions between ``matter'' and ``radiation'', is conformal.
For example, all particles can radiate gravitons, we just have to
appropriately understand the corresponding divergent amplitudes as
the generalized functions. With the similar reasoning one can
arrive at the conclusion that eternally accelerating charged
particle in Minkowski space (e.g. under the action of the constant
electric field) does emit radiation.

As the side remark let us point out here that one may object the
conclusions made in the previous paragraph based on the following
considerations. It is known that if one drops a spherically
symmetric massive ($m$) body in de Sitter space it produces the
static de Sitter black hole metric outside its own volume:

\bqa ds^2 = - \left(1 - r^2 - \frac{2\,m}{r^{D-3}}\right)\, dt^2 +
\frac{dr^2}{\left(1 - r^2 - \frac{2\,m}{r^{D-3}}\right)} \nonumber
\\ + r^2 \, d\Omega^2_{D-2} \label{BH}\eqa and does not produce any
(non--static) gravitational waves on top of that. Hence, it seems
that this argument precludes our conclusion that an inertial
massive body in de Sitter will produce gravitational waves. But
the important point is that such a metric as \eq{BH} is seen by a
non--inertial observer which is {\it fixed} above the surface of
the spherical body, i.e. the body and the observer compose a bound
state and do not move with respect to each other. Thus, it is not
an occasion that such an observer does not see any radiation from
the massive body. At the same time our statement is that {\it it
is} the inertial observer which sees the radiation from free
floating bodies in de Sitter space.

Let us see now what happens with the cross--section of a hard
process containing the radiation of the soft $\psi$ quantum by
$\Psi$ in de Sitter space. Because the amplitude \eq{ampmasshel}
is non--vanishing on--shell the virtuality of the $\Psi$ particle
is {\it not} related to the mass--shell momentum ($j_3$) of the
radiated $\psi$ particle. Indeed, to obtain the multiplicative
factor in the amplitude of a hard process, corresponding to the
radiation of the soft quantum by the hard one, we have to multiply
by $1/\lambda$ the same amplitude as \eq{ampmasshel} with the only
change of $\mu_1$ in the second wave function under the integral
in \eq{ampmasshel} by $\sqrt{M^2 + \lambda -
\left(\frac{D-1}{2}\right)^2}$. Obviously the amplitude is {\it
not} singular as $j_3 \to 0$, because $\lambda$ is not related to
$j_3$. In fact, we do not have $\delta$--function (imposing energy
conservation) which fixes the value of $\lambda$ as it was in
Minkowski space. Hence, we just have to integrate over all
possible values of the virtuality in the amplitude, unlike the
Minkowski space case. Thus, the cross--section is not divergent as
well, which can be explicitly seen via similar reasoning to the
one presented after eq. \eq{ampmasshel}. But even if it was
divergent, we would not have had the factorization of the
divergences due to such a behavior of $\lambda$. The latter fact
anyway spoils the cancellation of the IR divergences at {\it all
orders} \cite{Smilga:1985hp}.

At the same time it happens that loop diagrams in de Sitter space
do have IR divergences even for massive particles. Consider the
one loop self--energy diagram for the $\Psi$ particle. It has the
contributions of the form:

\begin{widetext}
\bqa \delta \Sigma \propto \int_{-\infty}^{+\infty} dt_1 \,
\int_{-\infty}^{+\infty}  dt_2 \int d\Omega_1 \int d
\Omega_2 \cosh^{D-1}(t_1) \, \cosh^{D-1}(t_2) \times \nonumber \\
\times F\left(\frac{D-1}{2} + i\, \mu_1, \frac{D-1}{2} - i\,
\mu_1; \frac{D}{2}; \frac{1 \pm Z}{2} \right)\times \nonumber \\
\times F\left(\frac{D-1}{2} + i\, \mu_2, \frac{D-1}{2} - i\,
\mu_2; \frac{D}{2}; \frac{1 \pm Z}{2} \right)\label{selfen} \eqa
\end{widetext}
where $Z$ is given by \eq{subst} and we have borrowed propagators
from \eq{greenfunc}. It is straightforward to see (using
\eq{propasymp}) that such an integral divergent in the IR, i.e. as
$Z(z,z')\to\infty$ (see e.g. similar discussion in
\cite{Polyakov:2007mm}). Indeed, if say $t_1\to-\infty$, while
$t_2\to+\infty$, then according to \eq{subst}, $ Z(z,z') \to
e^{t_2 - t_1}\,(1 + \cos\Omega)/4$ and the integrand expressions
in \eq{selfen} behave as:

\bqa \left[c_1\, e^{- i\,\mu_1\,(t_2 - t_1)} + c_2\, e^{
i\,\mu_1\,(t_2 - t_1)}\right]\,
\nonumber \\
\left[c'_1\, e^{- i\,\mu_2\,(t_2 - t_1)} + c'_2\, e^{
i\,\mu_2\,(t_2 - t_1)}\right] \eqa Hence, the integrals over $t_1,
\,t_2$ in \eq{selfen} are divergent if $\mu_2$ is pure imaginary,
i.e. when $m< (D-1)/2$. Furthermore, we have the IR divergence in
the causally connected region, because any points with $t \to
-\infty$ and $t\to +\infty$ are causally connected in de Sitter
space. Hence, restricting oneself to the region within the
cosmological horizon does not help to cut or get rid of such IR
divergences.

Thus, for any $M>0$, but $0<m<(D-1)/2$, we obtain finite tree
level contributions to cross--sections and there is nothing which
can cancel the IR divergences in the loop diagrams. It is probably
worth pointing out here that we can not interpret the divergence
in \eq{selfen} as an analog of the collinear one
\cite{Weinberg:1965nx} for many obvious reasons. At least it is
present for any value of the mass $M$ of the particle emitting the
radiation.

All the considerations above make us to conclude that the
evolution operator leading to such a diagram technic is not
unitary and the system of de Sitter background plus QFT is not
closed if the cosmological constant is held fixed. It is important
to stress that in the circumstances under consideration one can
not make the IR divergences to cancel by a unitary change of the
basis of the creation--annihilation operators.

\section{Conclusions}

 Before drawing any conclusions let us make a few side remarks about the divergences
of \eq{ampmasshel} when either of the masses $M$ or $m$ is
vanishing. Similar divergences appear in anti--de--Sitter space:
note the similarity of the anti--de--Sitter metric $ ds^2 =
\frac{1}{z^2}\,\left(dz^2 + dx_a\,dx^a\right)$ to the one in
\eq{planar} with the crucial difference, however, that $z$ is not
time--like. Hence, anti--de--Sitter space is not globally
hyperbolic (because there is time--like boundary) but does not
have an event horizon (because there is a globally defined
time--like Killing vector). Because of the lack of the global
hyperbolicity, which results in such a well known effect that in
anti--de--Sitter space waves can repel from the spacial boundary,
one can not define Cauchy problem in such a space
\cite{Hawking:1973uf}. Thus, while it is possible to define the
unique anti--de--Sitter invariant vacuum state, one can {\it not}
define appropriately the evolution operator for a QFT in such a
background. According to the AdS/CFT--correspondence
\cite{Maldacena:1997re}, \cite{Gubser:1998bc},
\cite{Witten:1998qj} one treats the IR divergences in the QFT on
anti--de--Sitter space, which appear in the wave functional rather
than in the S--matrix, as the UV divergences in the QFT on its
boundary.

On the contrary, although we do not have Poincare invariance in de
Sitter space, it {\it is} globally hyperbolic and one can define
evolution operator there. Let us stress here that we do not have
Poincare invariance, for example, in the presence of the constant
electric field in QED, but we still can define the evolution
operator, because the evolution problem can be correctly
formulated in such circumstances.

At this point we can and should address the question why
anti--de--Sitter space is stable? It seems that a free floating
particle in anti--de--Sitter space will emit radiation as well.
However, this question and the question of the cancellation of the
IR divergences can not be formulated in anti--de--Sitter space
because of the impossibility to define the time evolution operator
in such a background due to the lack of the global hyperbolicity.

Let us now come back to the conclusions. We see that QFT in de
Sitter space (i.e. if we fix the cosmological constant) behaves as
if it is formulated in a background of an external
quasi--classical gravitational field (excitation above the correct
vacuum) due to cosmological constant
--- analogously to the QED in a constant (in space and time) electric field,
i.e. as the non--closed system. That is the only interpretation of
the non--unitarity, which we can give. Hence, the only conclusion
which we can make here, without performing a direct calculation,
is that the cosmological constant will relax to zero creating
particles via Gibbons--Hawking pair production (which we do not
discuss in our paper) and via performing work on accelerating
created particles, which leads to the radiation in its own right.

The conceptual question is how fast will be the relaxation of the
cosmological constant? If the relaxation goes fast enough we can
hope to explain via this mechanism the cosmological constant
problem along with obtaining natural inflation without any
inflaton field \cite{Tsamis:1996qq}. We think that the rate of the
decay should just depend on the actual magnitude of the field. For
the big enough field the rate of the decay should be big: bigger
the energy pool is --- easier to create light particles. The
technical question is how to calculate the rate of the decay of
the cosmological constant?

If the cosmological constant is very big we have to use theory of
quantum gravity. Unfortunately string theory does not have a
formulation on de Sitter space, because such a background spoils
conformal invariance of the theory on the string worlds--sheet. In
this case we have to deal with off--shell formulation of the
closed string theory, which is very hard to do with the presently
existing first quantized variant of the theory (see, however,
\cite{Polyakov:2007mm}).

Thus, the only hope is that we can use the ordinary
Einstein--Hilbert theory, which dominates in the IR, if the
cosmological constant is smaller than the Plank scale. In this
respect we should stress that the issue of the instability of de
Sitter space, of the IR divergences and of the non--unitarity have
been discussed in various places \cite{Myrhvold},
\cite{Tsamis:1996qq}, \cite{Mottola:1984ar},
\cite{Polyakov:2007mm}, \cite{Antoniadis:1991fa},
\cite{Ford:1984hs}. As well the running of the cosmological
constant due to the quantum fluctuations have been found in
\cite{Dolgov:1994cq}, \cite{Tsamis:1996qq} and
\cite{Weinberg:2006ac} using either Heisenberg or
Schwinger--Keldish technics. Here, apart from presenting the
listed above new phenomena supporting the conclusion that de
Sitter space is unstable, we should criticize the actual
calculation of the decay rate of the cosmological constant
performed in \cite{Dolgov:1994cq}, \cite{Tsamis:1996qq} and
\cite{Weinberg:2006ac}, because, as we just pointed out, the rate
was found with the use of the non--unitary evolution operator.
Hence, the questions posed above remain to be answered.

\begin{acknowledgments}
Authors would like to thank V.Zakharov for collaboration at the
initial stage of work on this project. AET would like to thank M.
Voloshin, L. Kofman, V. Lukash, A. Mironov, A. Dolgov, V. Rubakov,
N. Nekrasov, M. Polikarpov, S. Theisen, P.Arseev, B.Voronov,
I.Polyubin and especially A. Rosly and O. Kancheli for valuable
discussions. AET would like to thank MPI--AEI, Golm, Germany for
the hospitality during the final period of work on this project.
As well authors would like to thank Galileo Galilei Institute for
Theoretical Physics for the hospitality and the INFN for the
partial support during the completion of this work. The work was
partially supported by the Federal Agency of Atomic Energy of
Russian Federation and by the grant for scientific schools
NSh-679.2008.2.
\end{acknowledgments}

\end{document}